\documentclass[prl,twocolumn,showpacs,floatfix,amsmath,amssymb,superscriptaddress] {revtex4}
\usepackage[dvips]{graphicx}
\usepackage{latexsym}
\usepackage{graphicx}
\usepackage{times}
\usepackage{amsmath}
\usepackage{dcolumn}
\usepackage{subfigure}
\usepackage{latexsym,amsmath,amssymb,bm,euscript}
\def\beq{\begin{equation}}
\def\eeq{\end{equation}}
\def\bea{\begin{eqnarray}}
\def\eea{\end{eqnarray}}

\begin{document}

\title{Bismuth in strong magnetic fields: unconventional Zeeman coupling and correlation effects}

\author{Jason Alicea}
\affiliation{Department of Physics, California Institute of  
Technology, Pasadena, CA 91125}
\author{Leon Balents}
\affiliation{Kavli Insitute for Theoretical Physics, University of California, Santa Barbara, CA 93106-9530}

\date{\today}

\begin{abstract}

  Recent experiments on bismuth have uncovered remarkably rich magnetization structure at fields well beyond the regime in which all carriers are expected to reside in the lowest Landau level.  Motivated by these findings, we start from a microscopic tight-binding model and derive a low-energy Hamiltonian for the holes and three Dirac electrons pockets in bismuth.  We find that an unconventional electron Zeeman effect, overlooked previously, suppresses the quantum limit for the electrons dramatically, giving rise to the observed anomalous magnetization structure.  We further study interaction effects near fields at which the 2nd Landau level for one electron pocket empties, where magnetization hysteresis was observed.  Here we find instabilities towards both charge density wave and Wigner crystal phases, and propose that hysteresis arises from a first-order transition out of the latter.

\end{abstract}
\pacs{}

\maketitle

\emph{Introduction}.  After more than a century of active research, bismuth continues to yield fascinating discoveries.  Much of this material's exceptional behavior stems from its band structure \cite{BandStructureReview}---the Fermi surface arises from a hole pocket and three Dirac electron pockets which contribute an extremely low carrier density of $3\times 10^{18}$cm$^{-3}$.  One remarkable consequence of the small carrier concentration is that weak Sb doping is believed to change the system from a semimetal to a topological insulator \cite{FuKane,TIexpt}. Another is that the quantum limit, at which carriers are confined to the lowest Landau level (LLL), can be realized with laboratory fields.  In particular, for fields along the highest-symmetry `trigonal' axis, the hole quantum limit occurs at $\sim$9T \cite{gfactorsOld}.  The electron quantum limit, while less clear experimentally, is believed to occur at similar fields \cite{Ong}.  

From a single-particle perspective, quantum oscillations should subside once all carriers reside in the LLL, and transport and thermodynamic quantities should appear `featureless'.  Recent experiments on bismuth in trigonal fields nevertheless observed surprisingly rich physics extending well beyond 9T.  Nernst peaks were resolved at 13.3, 22.3, and 30.8T, with the Hall resistance exhibiting step-like features in between, prompting the suggestion that fractionalization may be occurring in this 3D system \cite{Behnia}.  More recent torque magnetometry studies at fields near the trigonal axis additionally measured unanticipated magnetization structure---including hysteresis---persisting up to the largest fields probed (31T), providing further evidence for correlation physics \cite{Ong}.

\begin{figure}
\centering{
  \subfigure{\includegraphics[width=1.8in]{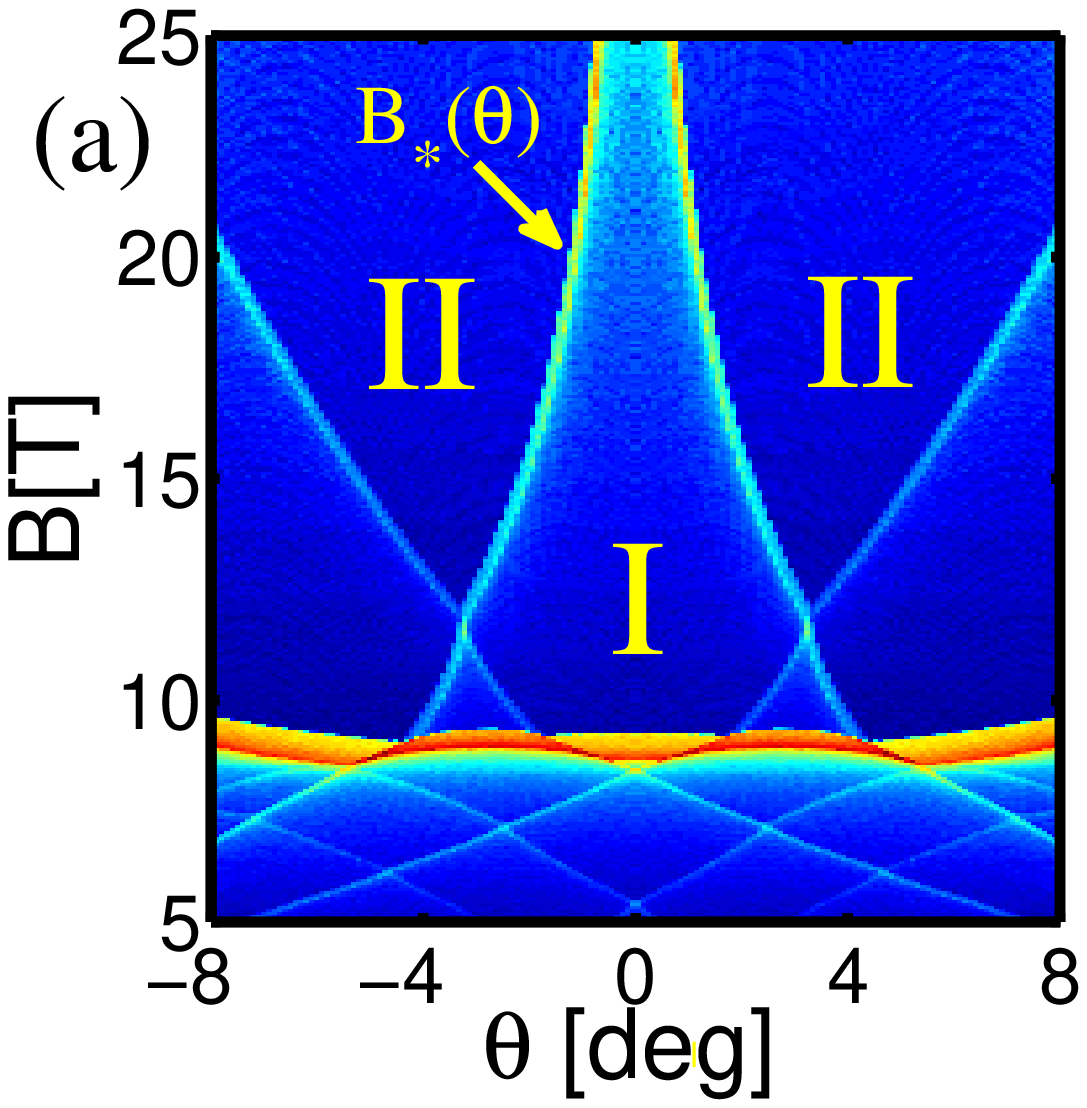}\label{DOS}}
  \subfigure{\includegraphics[width=1.57in]{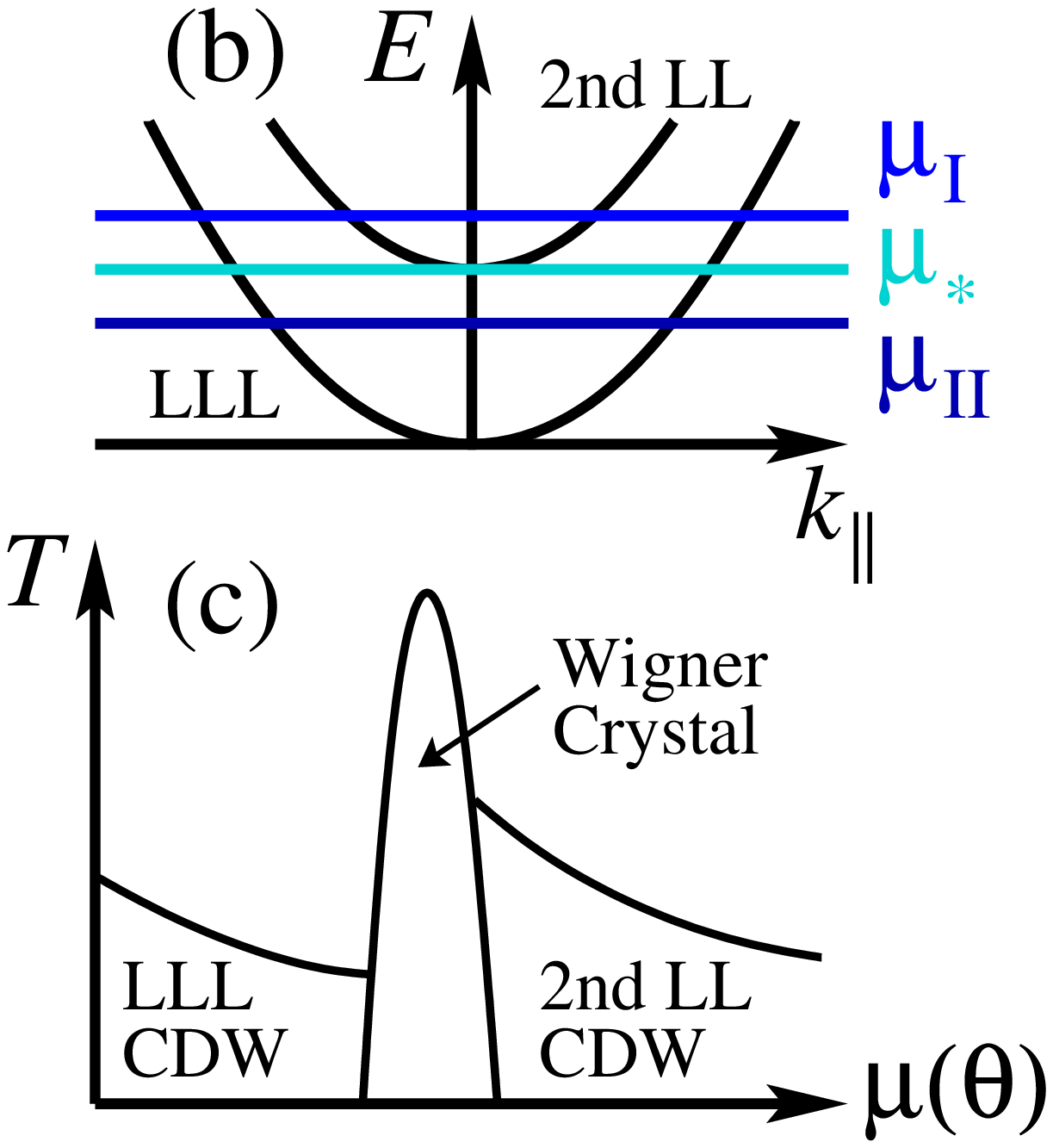}}
  \caption{(a) Single-particle DOS, excluding the contribution from the electron pocket invisible in torque experiments \cite{Ong}.  All electron pockets occupy the 2nd LL in region I, while pocket 3 empties into the LLL in region II.  (b) Schematic energy dispersion for pocket 3 and (c) proposed phase diagram with interactions near $B_*(\theta)$. }
  \label{DOSfigure}}
\end{figure}

Here we derive a low-energy theory for bismuth,
suitable for analyzing the Landau level (LL) structure and interaction effects.  For the electrons, 
we show that strong spin-orbit coupling generates unconventional Zeeman terms not present in the standard Cohen-Blount model \cite{CohenBlount}.  This coupling suppresses the
electron quantum limit far beyond 9T for trigonal fields
contrary to what has been assumed, recovering the high-field magnetization
structure observed in Ref.\ \onlinecite{Ong} [see Fig.\
\ref{DOSfigure}(a)].  Turning to Coulomb effects, we show
that interactions have the strongest influence near
fields for which a low LL empties.  In agreement with this
expectation, the experimentally observed hysteresis
coincides with one electron pocket emptying its 2nd LL.  To
address this aspect of experiment, we study interaction effects near
this band emptying, employing Functional Renormalization Group (FRG)
techniques similar to Ref.\ \onlinecite{Yakovenko} and going beyond the early analysis of Abrikosov \cite{Abrikosov}.  As the Fermi energy
passes from above the 2nd LL into the LLL, we find that the leading
instability for these electrons involves charge-density-wave (CDW) order
in the 2nd LL, followed by Wigner crystal formation and LLL CDW
order [see Fig.\ \ref{DOSfigure}(c)].  We suggest that the hysteresis
originates from the Wigner crystal phase, and discuss experiments to
verify this proposal.  Finally, we comment briefly on the
Nernst and Hall effect puzzles, which remain unexplained by this
work.

\emph{Low-energy theory}.  Our starting point is the tight-binding model
of Liu and Allen \cite{LiuAllen}, which was constructed to accurately
reproduce bismuth's band structure near the Fermi level
\cite{BandStructureReview}.  We derive from this microscopic model an
effective low-energy Hamiltonian for the electrons and holes in a
magnetic field ${\bf B}$ by expanding the lattice fermion operators as
follows,
\begin{equation}
  f_{\mu j \alpha}({\bf r}) \sim e^{i {\bf Q}\cdot {\bf
      r}}\varphi^\beta_{\mu j \alpha}h^\dagger_\beta + \sum_{\lambda =
    1}^3 e^{i {\bf P}_\lambda \cdot {\bf r}}\sum_{\ell = 1}^4
  \phi^{\lambda\ell}_{\mu j \alpha}\psi_{\lambda \ell}, 
  \label{expansion}
\end{equation}
where $\mu = s,p_{x,y,z}$ labels the outer-shell orbital with spin
$\alpha = \uparrow/\downarrow$ at site ${\bf r}$ on sublattice $j =
1,2$.  The two-component operators $h$ describe hole excitations near
wavevector ${\bf Q}$ and the four-component Dirac fermions
$\psi_\lambda$ describe electron excitations near ${\bf P}_\lambda$; the
corresponding wavefunctions are $\varphi^\beta$ and $\phi^{\lambda
  \ell}$.  Using Eq.\ (\ref{expansion}), we obtain the non-interacting
Hamiltonian $H_0 = \int_{\bf r}(h^\dagger \mathcal{H}_h h +
\sum_\lambda\psi_\lambda^\dagger\mathcal{H}_{e}^\lambda\psi_\lambda)$,
\begin{eqnarray}
    \mathcal{H}_h &=& -\mu_h -\frac{D_x^2 + D_y^2}{2m_\perp}- \frac{D_z^2}{2m_z} -\frac{g_h \mu_B B^z \sigma^z}{2} 
  \label{Hh} \\
  \mathcal{H}_{e}^3 &=& -\mu_e + m \mu^z -i\sum_{j = x,y,z} v_j D_j\eta^j -\frac{\mu_B {\bf G}\cdot {\bf B}}{2}.
  \label{He3}
\end{eqnarray}
Here ${\bf D} = \nabla-i q {\bf A}$, with ${\bf B} = \nabla \times {\bf
  A}$ and $q = \pm e$ for the holes and electrons, respectively.  We
employ three sets of Pauli matrices $\sigma^j$, $\tau^j$, and $\mu^j$
above, and define $\eta^x = \mu^x(v_{1x}\tau^y + v_{2x} \tau^z)/v_x$ and
$\eta^{y,z} = (v_{1y,z}\mu^y + v_{2y,z}\tau^x \mu^x)/v_{y,z}$, where
$v_{1j}^2 + v_{2j}^2 = v_j^2$.  The Hamiltonians $\mathcal{H}_{e}^{1,2}$
for pockets 1 and 2 can be obtained from $\mathcal{H}_{e}^3$ by rotating
${\bf D},{\bf B}$ by $\pm 2\pi/3$ about the trigonal ($z$) axis.  In
terms of the electron mass $m_e$ and speed of light $c$, we have $\mu_e = 0.0335$eV, $\mu_h = 0.012$eV, 
$m_\perp = 0.0675m_e$, $m_z = 0.612 m_e$, $m = 6$meV, $v_{1x} = 0.0022
c$, $v_{2x} = -0.002 c$, $v_{1y} = -7.6 \times 10^{-5}c$, $v_{2y} = 3.4
\times 10^{-4}c$, $v_{1z} = 0.002c$, $v_{2z} = -0.0014c$.  The Fermi energy $E_F = 0$ when ${\bf B = 0}$, but changes to maintain charge
neutrality when ${\bf B \neq 0}$.

Zeeman coupling warrants further discussion.  Microscopically, 
Zeeman energy has one source $\propto {\bf B}\cdot({\bf L} + 2{\bf
  S})$ and another coming from spin-orbit coupling $\propto (\nabla U
\times {\bf A})\cdot {\bf S}$, where $U$ is the crystal potential and
${\bf L}, {\bf S}$ denote orbital/spin angular momentum.  While the
corresponding low-energy terms can be obtained via Eq.\
(\ref{expansion}), evaluating the spin-orbit contribution is nontrivial
since the potential $U$ is unknown.  Progress can be made, however, by
assuming $U$ is rotationally invariant, as then only two independent
matrix elements remain: $\chi_{s/p} = \frac{1}{4m_e c^2}\langle s/p_x
\uparrow|r\partial_r U \sin^2\theta|s/p_x\uparrow\rangle$, where
$|s/p_x\uparrow\rangle$ are the atomic $s/p_x$ orbital wavefunctions.
With this assumption, we obtain the hole Zeeman splitting in Eq.\
(\ref{Hh}), which is sensitive only to $B^z$, consistent with experiment
\cite{gfactorsOld}.  For the electrons, we obtain the Zeeman coupling in
Eq.\ (\ref{He3}) with $G_x = \tau^x(g_{1x} + g_{2x}\mu^z)$ and $G_{y,z}
= \tau^y(g_{1y,z} + g_{2y,z}\mu^z) + \tau^z(g_{3y,z} + g_{4y,z}\mu^z)$.
The $g$-factors are listed in Table \ref{parameters}.  Crucially, $g_h$
depends only on $\chi_p$, and the electron $g$-factors obtain only a
weak contribution from $\chi_s$ because the wavefunctions $\phi^{\lambda
  \ell}$ are concentrated on the $p$-orbitals.  The precise value of
$\chi_s$ is therefore unimportant, and we will simply set $\chi_s =
\chi_p$.  Finally, hole Zeeman splitting has been well-studied
experimentally \cite{holegfactor}, and from the available data we deduce
that $g_h \approx 54$, which allows us to determine $\chi_p$ and hence
the electron $g$-factors.\footnote{The data of Ref.\
  \onlinecite{holegfactor} appear consistent with the hole Zeeman
  splitting being $2\pm 0.16$ times the cyclotron energy.  We use the
  lower sign, which gives $g_h = 54$, since the hole quantum limit then
  occurs closer to 9T as observed experimentally.  }

\begin{table}[h]
\begin{center}
\begin{tabular}{|c|c|c|c|c|c|c|c|}
\hline
 & $a$ & $b$ & $c$ & & $a$ & $b$ & $c$\\
\hline
$g_h$ & 4 & 2 & 0 & $g_{4y}$ & $-0.097$ & $-0.35$ & $-0.069$ \\
\hline
$g_{1x}$ & 1.5 & 1.3 & 0.072 & $g_{1z}$ & 0.32 & 0.18 & $-0.024$ \\
\hline
$g_{2x}$ & $-1.4$ & $-0.44$ & $-0.024$ & $g_{2z}$ & 0.68 & 0.38 & 0.07\\
\hline
$g_{1y}$ & $-1.5$ & $-0.9$ & $-0.019$ & $g_{3z}$ & $-0.18$ & $-0.42$ & $-0.019$ \\
\hline
$g_{2y}$ & $-1.3$ & $-0.76$ & 0.0037 & $g_{4z}$ & $-0.017$ & 0.26 & 0.0038 \\
\hline
$g_{3y}$ & 0.17 & 0.39 & 0.023 & & & & \\
\hline
\end{tabular}
\caption{\label{parameters} Electron/hole $g$-factors, with $g_\alpha = a_\alpha + b_\alpha \chi_p + c_\alpha \chi_s$.  }
\end{center}
\end{table}

Equations (\ref{Hh}) and (\ref{He3}) constitute the first major result
of this paper.  Most importantly, the electron Zeeman coupling has not
been considered previously, and modifies the spectrum dramatically at high fields as we now discuss.

\emph{Single-particle spectrum}.  While the hole Hamiltonian is
easily diagonalized, the electron part is much
more difficult since the Zeeman and orbital terms do not commute.
Qualitatively, Zeeman coupling generates components of higher LLs in the wavefunctions compared to the orbital-only problem and splits the usual LL degeneracy in the Dirac spectrum (see, \emph{e.g.}, Supporting Online Material for Ref.\ \onlinecite{Ong}).  To
proceed, we truncate the Hilbert space,
including only the first $n \sim 10$ LLs, and diagonalize the Hamiltonian
numerically.

Motivated by magnetization experiments \cite{Ong}, we have studied the
spectrum for fields tilted by an angle $\theta$ from the trigonal
towards the binary axis.  Figure \ref{DOS} displays the single-particle
density of states (DOS) in the $B-\theta$ plane (excluding the electron
pocket invisible to experiments \cite{Ong}).  Bright lines occur where
the Fermi energy crosses the bottom of a LL, and the flat line at $\sim$9T corresponds to the hole quantum limit.
Remarkably, the electrons give rise to features at much higher fields
which agree well with the anomalous structure reported experimentally
(see Fig.\ 3 in Ref.\ \onlinecite{Ong}).

The persistence of this structure up to such large fields arises from an increase in carrier density with field \cite{gfactorsOld} and the splitting of the electron LL degeneracy by Zeeman coupling, which makes 2nd LL states available at lower energies and strongly suppresses the electron quantum limit.  Indeed, in region I of Fig.\ \ref{DOS} all electron pockets occupy the 2nd LL.  
Tilting the field into region II pushes one of those pockets (assumed to be pocket 3 hereafter) into the LLL.  The dispersion for pocket 3 versus the momentum $k_\parallel$ along the field appears schematically in Fig.\ \ref{DOSfigure}(b), together with the chemical potential in I and II.  Interestingly, experiments observed hysteresis in the magnetization at the line labeled $B_*(\theta)$ separating these regions [corresponding to $\mu_*$ in Fig.\ \ref{DOSfigure}(b)].  Addressing this puzzle requires moving beyond single-particle physics.

\emph{Correlation effects}.  We now add Coulomb interactions,
\begin{equation}
  H_{\rm int} = \frac{1}{2}\int_{\bf r r'}V({\bf r-r'})\rho({\bf r})\rho({\bf r'}),
\end{equation}
where $\rho = \sum_\lambda\psi^\dagger_\lambda\psi_\lambda-h^\dagger h$ and $V({\bf r})$ is the screened Coulomb potential.  Note that the number of holes and electrons within each pocket remains separately conserved here.  Short-range pieces arising from lattice effects break this symmetry, but are subdominant and can be neglected.  

Interactions can be most simply treated at weak coupling, which
is controlled provided the dimensionless interaction strengths satisfy
$e^2/v_a^n \ll 1$, where $v_a^n$ is the Fermi velocity for pocket $a$ in
LL $n$.  This criterion inevitably breaks down near a band
emptying since at least one $v_a^n \rightarrow 0$.  Correlation effects will be most pronounced in this strongly coupled regime, and this is indeed where hysteresis appears.  Our aim now is to study the crossover from weak to
strong coupling to understand this aspect of experiment.  Rather than addressing the fate of all carriers, we will focus more narrowly on the \emph{leading} instability involving pocket 3 electrons near $B_*(\theta)$.

Even this restricted question requires considering numerous interactions, since $H_{\rm int}$ couples pocket 3 to all other
carriers, some of which occupy more than one LL.  We can, however,
further distill the problem using an intuitive principle: similar
carriers generally couple more effectively than dissimilar ones.  For
instance, velocity anisotropies sharply distort the electron LL wavefunctions, which strongly suppresses instabilities involving \emph{different} electron pockets.\footnote{Fields oriented so that 2 or 3
pockets are symmetry-related may produce instabilities involving
multiple electron pockets.}  Furthermore, the hole and electron Fermi
velocities differ significantly, which disfavors
electron-hole pairing.  We
will therefore concentrate on interactions involving \emph{only} pocket
3 electrons, commenting briefly on other couplings.  For the
three important cases [regions I, II, and the line $B_*(\theta)$], we
employ the FRG method to determine the leading instability.  FRG
equations give the renormalized interactions at length scale $L$ as a
function of the logarithmic rescaling factor $\ell = \ln (L/\lambda)$, where
$\lambda$ is a microscopic length of order the Fermi wavelength.

We begin in region II, where pocket 3 is confined to the LLL.  To obtain a low-energy Hamiltonian we linearize the kinetic energy around the Fermi momenta $\pm k_{F3}^0$, expand $\psi_3$ in terms of right/left movers, and project onto the LLL (we employ Landau gauge and label the transverse momentum $k_\perp$).  Interactions between pocket 3 electrons can then be written
\begin{eqnarray}
  H_{\rm int}^{3,{\rm II}} &=& \int_{{\bf k}_i}\rho(k_{\perp 1},k_{\perp 2})c_{R3}^{0\dagger}({\bf k}_1 + {\bf k}_3)  c^{0\dagger}_{L3}({\bf k}_2 + {\bf k}_3)
  \nonumber \\
  &\times& c_{L3}^0({\bf k}_3) c_{R3}^0({\bf k}_1+{\bf k}_2+{\bf k}_3),
\end{eqnarray}
where ${\bf k} = (k_\perp,k_\parallel)$ and $c^{0\dagger}_{R/L3}$
creates a right/left-moving LLL electron.  The FRG equation for the
coupling function $\rho(k,x) \equiv \rho({\bf r})$ Fourier transformed
in the second argument is
\begin{eqnarray}
  \partial_\ell \rho({\bf r}) = \frac{1}{2\pi v_3^0}[\rho({\bf r})^2 - (\rho*\rho)({\bf r})],
  \label{LLLRG}
\end{eqnarray}
with $(f*g)({\bf r}) = \int_{{\bf r}'{\bf r}''}e^{i[{\bf r}\wedge{\bf
    r}'+{\bf r}'\wedge{\bf r}''+{\bf r}''\wedge{\bf r}]}f({\bf
  r}')g({\bf r}'')$.  The solution to Eq.\ (\ref{LLLRG}) is
well-understood \cite{Yakovenko,Tsai}.  The first term in the $\beta$
function causes $\rho$ to flow off, driving condensation of $\langle
c_{3R}^{0\dagger}c_{3L}^0\rangle$ which gaps this channel and yields
$2k_{F3}^0$ CDW order along the field; the second
term merely reduces the transition temperature \cite{Yakovenko}.
Furthermore, it follows from the initial conditions that $\rho({\bf r =
  0})$ is largest at the instability, so that the transverse density is
uniform in the CDW phase.  As an aside, we note that coupling to other
electron pockets and the holes leads to a series of couplings with flow
equations analogous to Eq.\ (\ref{LLLRG}).  Due to the velocity
anisotropy discussed above, inter-pocket electron instabilities set in
at very low energies, and are preempted by the $\rho$ instability.
Similarly, since the hole velocity is several times smaller than
$v_3^0$, hole-hole pairing preempts the `excitonic' instability \cite{HalperinRice} in the electron-hole channel.

Next, we sit at $B_*(\theta)$ and fine-tune the chemical potential to
$\mu_*$ in Fig.\ \ref{DOSfigure}(b), precisely at the bottom of the 2nd
LL.  At low energies, we now have linearly-dispersing right/left movers
$c_{R/L3}^0$ coupled with \emph{quadratically} dispersing 2nd LL
electrons which we denote by $d_3^1$.  As it stands, the problem can not
be treated within weak coupling since the soft dispersion for the latter
renders interactions involving $d_3^1$ strongly relevant at the
non-interacting fixed point.  To proceed, we follow Ref.\
\onlinecite{2chainHubbard} and seek a controlled limit by replacing the
2nd LL kinetic energy $Dk_\parallel^2$ with
$D|k_\parallel|^{1+\epsilon}$ and performing an expansion in $\epsilon$
\emph{and} the interaction strength in the limit $1 \gg \epsilon \sim
e^2/v_3^0$.

Pocket 3 interactions now involve several couplings:
\begin{eqnarray}
  H_{\rm int}^{3,B_*} &=&\int_{{\bf k}_i}\{\rho(k_{\perp 1},k_{\perp 2})c_{3R}^{0\dagger}({\bf k}_1 + {\bf k}_3)  c^{0\dagger}_{L3}({\bf k}_2 + {\bf k}_3)
  \nonumber \\
  && \!\!\!\!\!\!\!\!\!\!\!\!\!\!\!\!\!\!\!\!\!\!\! \times c_{L3}^0({\bf k}_3) c_{R3}^0({\bf k}_1+{\bf k}_2+{\bf k}_3) +u d_{3}^{1\dagger}d^{1\dagger}_{3}d_{3}^1d_{3}^1
   \\
  && \!\!\!\!\!\!\!\!\!\!\!\!\!\!\!\!\!\!\!\!\!\!\! + [v c_{3R}^{0\dagger}d^{1\dagger}_{3}d_{3}^1c_{R3}^0 + (R\rightarrow L)]+ [w c_{3R}^{0\dagger}c^{0\dagger}_{3L}d_{3}^1d_{3}^1 + h.c.]\}.
  \nonumber
\end{eqnarray}
The arguments in the last three couplings have been suppressed for brevity, but should appear as in the $\rho$ term.  
Fourier transforming in the second argument and defining ${\bf r} = (k,x)$ as before, it will be convenient below to write the flow equations for these coupling functions as follows:
\begin{eqnarray}
  \partial_\ell \rho({\bf r}) &=& \frac{1}{2\pi v_3^0}[\rho({\bf r})^2-(\rho*\rho)({\bf r})]-\frac{\alpha}{\pi}(w*w^*)({\bf r}),
  \nonumber \\
  \partial_\ell u({\bf r}) &=& \epsilon u({\bf r}) +\frac{\beta}{\pi}u({\bf r})^2-\frac{\alpha}{\pi}(u*u)({\bf r})
  \nonumber \\
  &-& \frac{1}{2\pi v_3^0}(w^* * w)({\bf r}),
  \label{BBRG} \\
  \partial_\ell v({\bf r}) &=& \frac{\gamma_1}{2\pi}[v({\bf r})^2-(v*v)({\bf r})] +  \frac{2\gamma_2}{\pi}|w({\bf r})|^2,
  \nonumber \\
  \partial_\ell w({\bf r}) &=& \frac{\epsilon}{2} w({\bf r}) -\frac{1}{2\pi v_3^0}(\rho*w)({\bf r})-\frac{\alpha}{\pi}(w*u)({\bf r})
  \nonumber \\
  &+& \frac{\gamma_1}{\pi}v({\bf r})w({\bf r}) -\frac{\delta}{\pi}\int_{\bf r'}e^{i{\bf r}\wedge {\bf r}'}v({\bf r}')w({\bf r'}).
  \nonumber
\end{eqnarray}
Here $\alpha = 1/D$, $\gamma_{1} = \delta = 1/(v_3^0+D\lambda^\epsilon e^{-\epsilon \ell})$, $\gamma_2 = \gamma_1 \lambda^\epsilon e^{-\epsilon\ell}$, and $\beta = 0$.  Naively, power-counting suggests that $u$ flows off first.  However, since $\beta = 0$, the analogous term that drove the CDW instability in region II is absent, which leaves open a more interesting possibility.  

To facilitate analytic and numeric progress, we now approximate $V({\bf
  r})$ as local (anisotropies can then be scaled away).  Note first that
at the initial conditions, $w$ is suppressed compared to $\rho, u,v$
since terms that survive Pauli exclusion are suppressed due to small
overlaps between the participating LL wavefunctions.  Thus we begin by setting $w = 0$, which decouples the remaining equations.  Clearly both $\rho$ and $v$ then flow off, with the former diverging faster since  $\rho$ is larger at the initial conditions.  The behavior of $u$ is less obvious.  Assuming $u(k,x)$ is
rotationally invariant (after rescaling) and analytic in $k,x$, we can
solve the flows for $u$ by assuming an ansatz $u({\bf r};\ell) =
\sum_jf_j(\ell)\chi_j(r)$.  Here $j$ runs over $0, 2, 4,\ldots$ and
$\chi_j(r) = e^{-r^2/2}P_j(r)$, with $P_j(r)$ degree-$j$ polynomials
defined so that $(\chi_j*\chi_{j'})({\bf r}) =
\delta_{j,j'}C_j\chi_j(r)$.  One then obtains decoupled equations for
each $f_j$ which show that $u({\bf r};\ell)$ flows to a finite fixed
point provided $C_jf_j(0) \geq 0$ for all $j$.  In our problem
the initial conditions are well-approximated by $u({\bf r}; \ell = 0)
= c(r^2-1)e^{-r^2/2}\propto \chi_2(r)$, with $c>0$, which indeed remains
finite under renormalization.  Though this result is correct in the
$\epsilon$-expansion, we  caution that the fixed point {\sl could} move
to strong coupling for physical $\epsilon=1$.  This would imply a
$u$-driven instability (see below).

To treat the full problem with $w \neq 0$ we rely on numerical
integration of the FRG equations.  Here we find that $w$ further
enhances the instability in $\rho$, and also generates additional
non-zero components of $\chi_j$ in $u$, causing this function to flow
off in tandem with $w$.  In contrast, $v$ is only weakly affected by $w$
and remains unimportant.  Because $\rho$ was already unstable when $w =
0$, this function becomes of order one first, driving CDW order in the
LLL with uniform transverse density.  The leading instability in the 2nd
LL occurs soon after driven by $u$, which localizes these states
producing Wigner crystallization as the chemical potential increases.
While this is certainly reasonable, we caution that definitively ruling
out the scenario where $u$ provides the leading instability is difficult
due numerical limitations, particularly when $V({\bf r})$ is non-local.
In this case the order of the instabilities is simply reversed.  

Finally, consider region I, with the chemical potential slightly above
the 2nd LL for pocket 3.  This problem is reminiscent of a 1D wire with
2 transverse modes occupied \cite{Meyer}, though the LLs
change the physics qualitatively.  As in region II, we derive a
low-energy theory for right/left movers in the LLL,
$c_{3R/L}^0$, and 2nd LL, $c_{3R/L}^1$.  Interactions then read
\begin{eqnarray}
  H_{\rm int}^{3,{\rm I}} &=& \int_{{\bf k}_i}\{\rho(k_{\perp 1},k_{\perp 2})c_{3R}^{0\dagger}({\bf k}_1 + {\bf k}_3)  c^{0\dagger}_{L3}({\bf k}_2 + {\bf k}_3)
  \nonumber \\
  && \!\!\!\!\!\!\!\!\!\!\!\!\!\!\!\!\!\!\!\!\!\!\! \times c_{L3}^0({\bf k}_3) c_{R3}^0({\bf k}_1+{\bf k}_2+{\bf k}_3) +u c_{3R}^{1\dagger}c^{1\dagger}_{3L}c_{3L}^1c_{3R}^1
   \\
  && \!\!\!\!\!\!\!\!\!\!\!\!\!\!\!\!\!\!\!\!\!\!\! + [v c_{3R}^{0\dagger}c^{1\dagger}_{3L}c_{3L}^1c_{R3}^0 + (R\rightarrow L)]+ [w c_{3R}^{0\dagger}c^{0\dagger}_{3L}c_{3L}^1c_{3R}^1 + h.c.]\},
  \nonumber
\end{eqnarray}
where again the arguments of the last three couplings should appear as
in the first.  These couplings flow as in Eqs.\ (\ref{BBRG}), but with
$\epsilon = \delta = 0$, $\alpha = \beta = 1/(2v_3^1)$, and $\gamma_1 =
4\gamma_2 = 2/(v_3^0+v_3^1)$.  It is natural to suspect here that $u$
dominates, since the 2nd LL carriers have the slowest
velocity.  This is indeed correct, which can be understood analytically
by ignoring all $(f*g)$ terms in Eqs.\ (\ref{BBRG}).  Since $\beta$ is
the largest coefficient remaining when the 2nd LL is weakly populated,
$u$ flows off before all other couplings, driving CDW order formed by
condensing $\langle c_{3R}^{1\dagger}c_{3L}^1\rangle$.  The transverse
density becomes Wigner crystalline \cite{Tsai} at the instability, since
the form of the 2nd LL Dirac wavefunctions dictates that $u({\bf r})$
is maximized at ${\bf r \neq 0}$.  In the limit where $V({\bf r})$ is
local, we have verified numerically that terms neglected in this crude
analysis do not modify these conclusions.
  
\emph{Discussion}. Putting our results together, we propose the minimal
phase diagram shown in Fig.\ \ref{DOSfigure}(c).  We speculate that
experiments of Ref.\ \onlinecite{Ong} may be conducted above the
critical temperature for the CDW phases in regions I and II but below that for the Wigner
crystal phase near the boundary.  The latter transitions are almost certainly first order,
which would be consistent with the observed hysteresis.  However,
\emph{two} transitions ought to occur here, whereas experiments see only
one.  This issue can be resolved by invoking disorder, which will be
particularly important in the low-density region of the Wigner crystal
phase close to region II.  The transition on that side is expected to be
smeared by disorder-pinning of the localized states, which 
should be addressed in more detail.  Future
experiments, particularly nonlinear transport and x-ray scattering
studies to search for signs of CDW and Wigner crystal order, should
provide valuable clues as to the true nature of this transition.
  
The Hall and Nernst effect puzzles in
trigonal fields $B > 9$T \cite{Behnia} are difficult to resolve at weak coupling.  Here hole-hole pairing drives the leading CDW instability, which is not
expected to recover these anomalies.  While surface states
should be seriously considered as emphasized in Ref.\ \onlinecite{Ong}, we
believe a more exotic origin is not inconceivable.
Interactions between the LLL holes are \emph{not} weak, as $e^2/v_h^0
\sim 24$, well outside of the range where weak coupling is expected to
be reliable.  In contrast, interactions between LLL electrons, while not
necessarily weak, are several times smaller: $e^2/v_e^0 \sim 5$.
Although screening should reduce these somewhat, the problem warrants
studying from a strong-coupling standpoint, which presents an exciting research direction \cite{Bernevig}.  
  
\acknowledgments{It is a pleasure to acknowledge illuminating
  discussions with K. Behnia, Lu Li, O. Motrunich, N.P. Ong, and
  G. Refael, as well as the hospitality of the KITP where this work was
  initiated.  We also acknowledge support from the Lee A.\ DuBridge
  Foundation (JA), and the Packard Foundation and National Science
  Foundation through grants DMR-0804564 and PHY05-51164 (LB).}


\end{document}